\documentstyle[11pt,mrs2001,epsfig]{article}
\def\hii{H{\sc ii}}
\def\mabs{$M_{\rm B}$}
\def\doh{$12 + \log(\rm O/H)$}
\begin{document}
\title{Chemical Properties of Starburst Galaxies Near 
and Far: Clues to Galaxy Evolution}

\author{T.\,Contini $^{1,2}$, M.-A.\,Treyer $^{3}$, M.\,Mouhcine $^{2}$, 
M.\,Sullivan $^{4}$, \& R.\,S.\,Ellis $^{5}$}
\vspace{0.2cm}
\affil{$^1$ Observatoire Midi-Pyr\'en\'ees, UMR 5572, 14 avenue E. Belin, F-31400 Toulouse, France}
\affil{$^2$ Observatoire de Strasbourg, 11 rue de l'Universit\'e, F-67000 Strasbourg, France}  
\affil{$^3$ Laboratoire d'Astrophysique de Marseille, Traverse du Siphon, F-13376 Marseille, France}
\affil{$^4$ Institute of Astronomy, Madingley Road, Cambridge, CB3 OHA, UK}
\affil{$^5$ California Institute of Technology, Pasadena, CA 91125, USA}

\begin{abstract}
The determination of chemical abundances in star-forming galaxies and 
the study of their evolution on cosmological timescales are powerful 
tools for understanding galaxy formation and evolution. This contribution 
presents the latest results in this domain. We show that detailed studies 
of chemical abundances in UV--selected, \hii\ and starburst nucleus galaxies, 
together with the development of new chemical evolution models, put 
strong constraints on the evolutionary stage of these objects in terms 
of star formation history. Finally, we summarize our current knowledge 
on the chemical properties of intermediate-- and high--redshift galaxies. 
Although the samples are still too small for statistical studies, 
these results give insight into the nature and evolution of distant 
star--forming objects and their link with present--day galaxies.
\end{abstract}

\section{Introduction}

Tracing the star formation history of galaxies is essential 
for understanding galaxy formation and evolution. The chemical 
properties of galaxies are closely related to their star formation 
history, and can be considered as fossil records, enabling us 
to track the galaxy formation history down to the present. 

There is a lot of observational evidence suggesting that 
the star formation history of galaxies has not been monotonic
with time, but exhibits instead significant fluctuations. 
Galaxies in the Local Group are excellent examples showing a 
variety of star formation histories. 
Further evidence for the ``multiple-burst'' scenario was recently 
found in massive Starburst Nucleus Galaxies (SNBGs; e.g. 
\cite{Cozetal99,Lancetal01}). 
Even the small-mass and less evolved \hii\ galaxies seem to be formed 
of age-composite stellar populations indicating successive bursts of 
star formation (e.g.\,\cite{Raietal00}). 

\cite{KCB01} recently explored numerical models of galaxy evolution in 
which star
formation occurs in two modes: a low-efficiency continuous mode, and 
a high-efficiency mode triggered by interaction with a
satellite. With these assumptions, the star formation history of
low-mass galaxies is characterized by intermittent bursts of star
formation separated by quiescent periods lasting several Gyrs, whereas
massive galaxies are perturbed on time scales of several hundred Myrs
and thus have fluctuating but relatively continuous star formation
histories. In these models, merger rates are specified using the
predictions of hierarchical galaxy formation models (e.g.\,\cite{Coleetal00}).

Examining the chemical evolution and physical nature of star-forming
galaxies over a range of redshifts will shed light on this issue.
Emission lines from \hii\ regions have long been the primary means of
chemical diagnosis in local galaxies, but this method has only
recently been applied to galaxies at cosmological distances following
the advent of infrared spectrographs on 8 to 10-m class telescopes
(e.g.\,\cite{Petetal98,KZ99,KK00,CL01,Hametal01,Petetal01}) .

\section{The case of UV-selected galaxies}

\cite{Contetal01} recently derived the chemical properties of 
a UV-selected sample of galaxies. These objects are found to be 
intermediate between low-mass, metal-poor \hii\ galaxies and more 
massive, metal-rich SBNGs (see \cite{Cozetal99} for the dichotomy), 
spanning a wide range of oxygen abundances, from $\sim$ 0.1 to 1 Z$_{\odot}$. 

\begin{figure}[t]
\plottwo{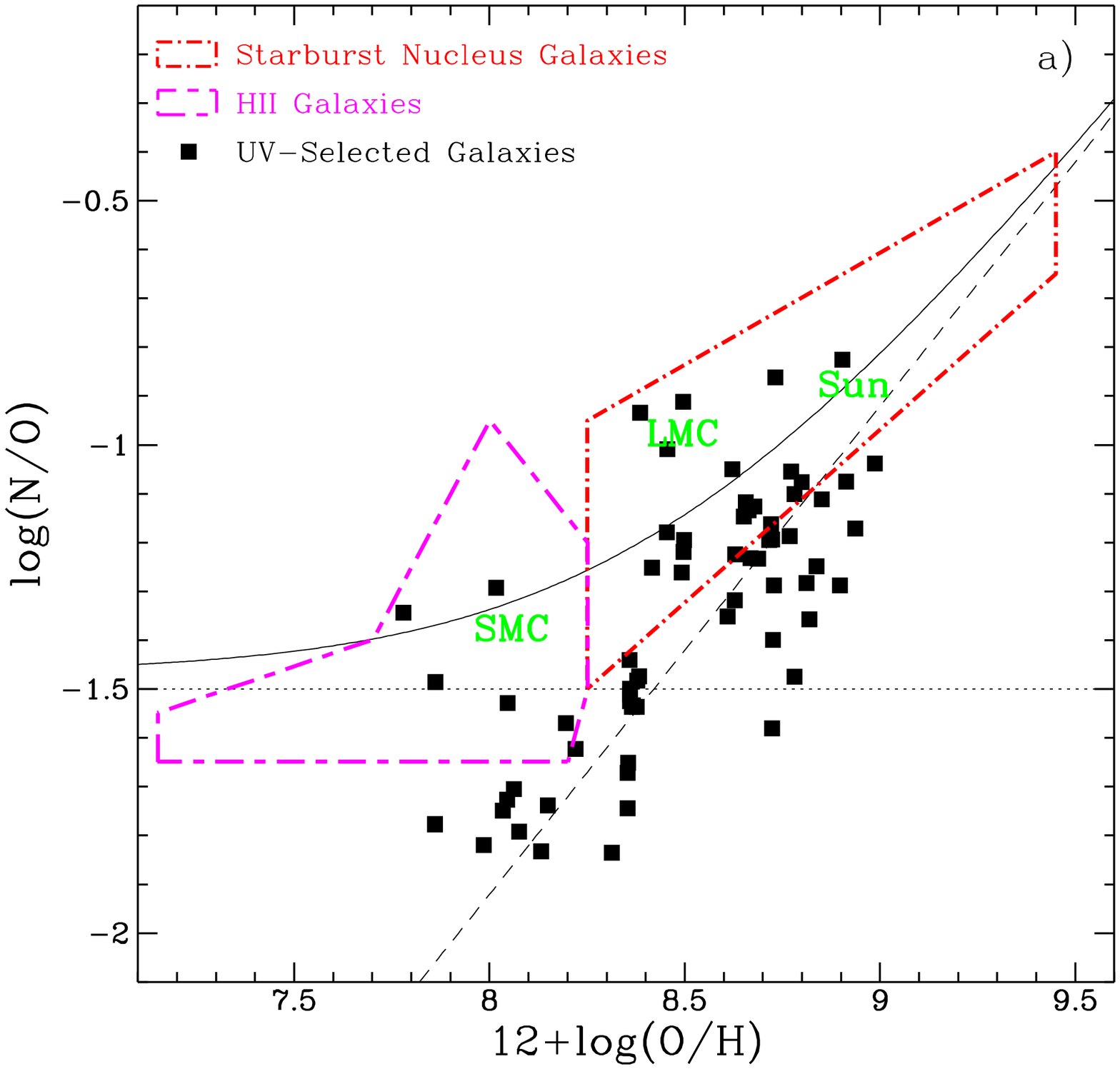}{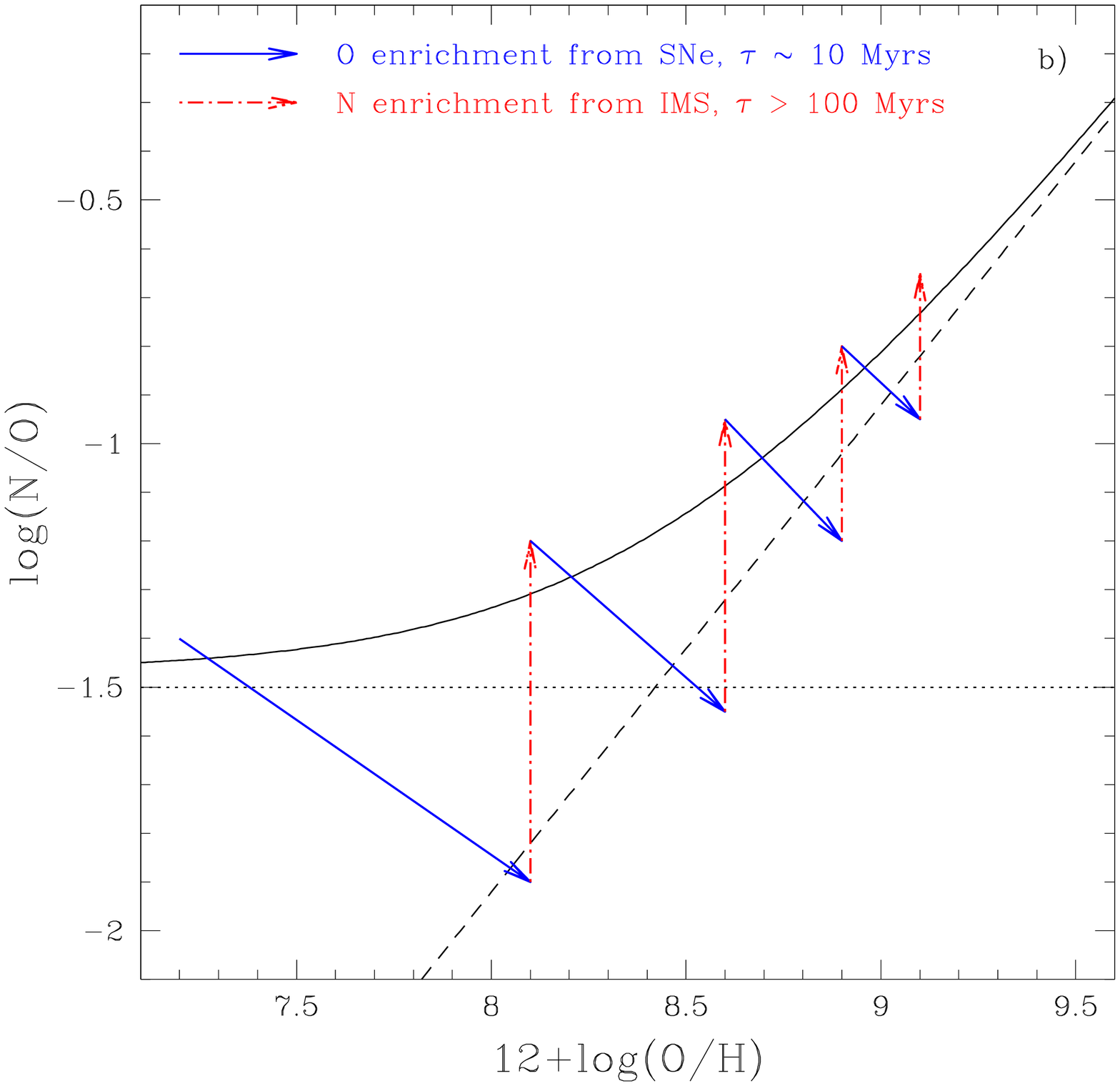}
\caption{{\em a}) N/O versus \doh\ for the
UV-selected galaxies (squares; \cite{Contetal01}) and two comparison
samples of nearby star-forming galaxies: Starburst Nucleus Galaxies 
(dot -- short dash line) selected in the optical (\cite{CCD98,Consetal00}) 
or in the far-infrared (\cite{Veietal95}), 
and \hii\ galaxies (short dash -- long dash line; \cite{KS96,IT99}). 
Theoritical
curves for a {\em primary} (dotted line), a {\em secondary} (dashed
line), and a {\em primary + secondary} (solid line) production of
nitrogen (\cite{VCE93}) are shown. Abundances ratios for the
Sun, the LMC and the SMC are also indicated.
{\em b}) Schematic evolution model of N/O versus
\doh\ assuming a sequence of starbursts separated by quiescent periods
(see also \cite{G90,Cozetal99}). This
scenario assumes that each burst first produces oxygen
enrichment due to massive star evolution on short time-scales
($\tau \sim 10$ Myrs), followed by significant nitrogen enrichment
on longer time-scales ($\tau > 100$ Myrs) due to intermediate-mass star
evolution (see \cite{Contetal01} for details).}
\label{UV}
\end{figure}

The behavior of these starburst galaxies in the N/O versus O/H 
relation (see Figure~\ref{UV}a) has been investigated in order to probe 
their physical 
nature and star formation history (see \cite{Contetal01} for details). 
At a given metallicity, the majority of UV-selected galaxies 
has low N/O abundance ratios whereas SBNGs show an excess of nitrogen 
abundance when compared to \hii\ regions in the disk of normal
galaxies (see also \cite{Cozetal99}). The interpretation of these 
behaviors is not straightforward. 
A possible interpretation of the location of UV-selected galaxies and
SBNGs in the N/O versus O/H relation could be that UV galaxies are selected 
at the end of a short episode of star formation following a rather 
long and quiescent period (\cite{Contetal01}), whereas SBNGs 
experienced successive starbursts over the last Gyrs to produce the 
observed nitrogen abundance excess (e.g.\,\cite{Cozetal99}) 

\section{Constraining the star formation history of galaxies}

\begin{figure}[t]
\plottwo{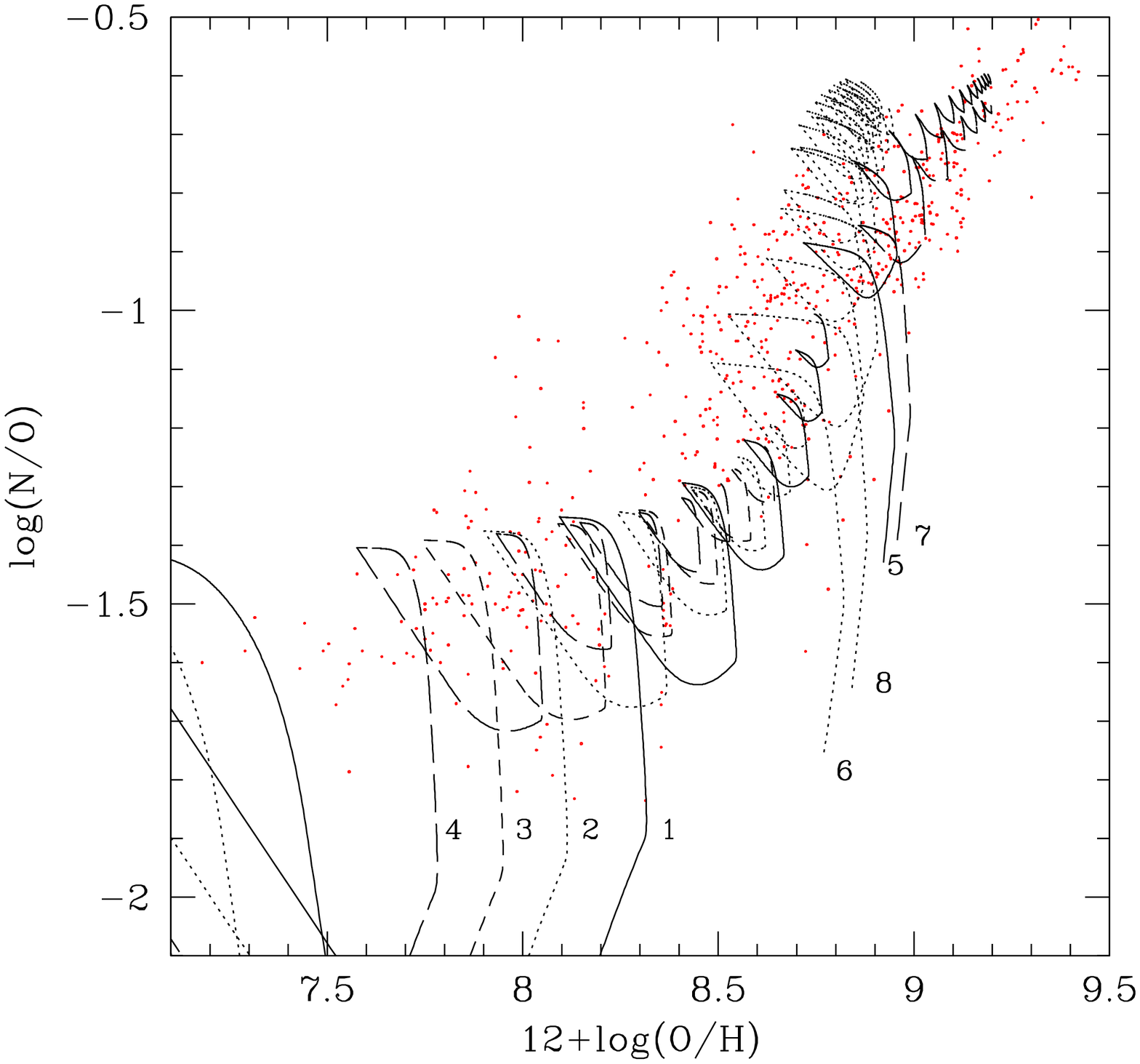}{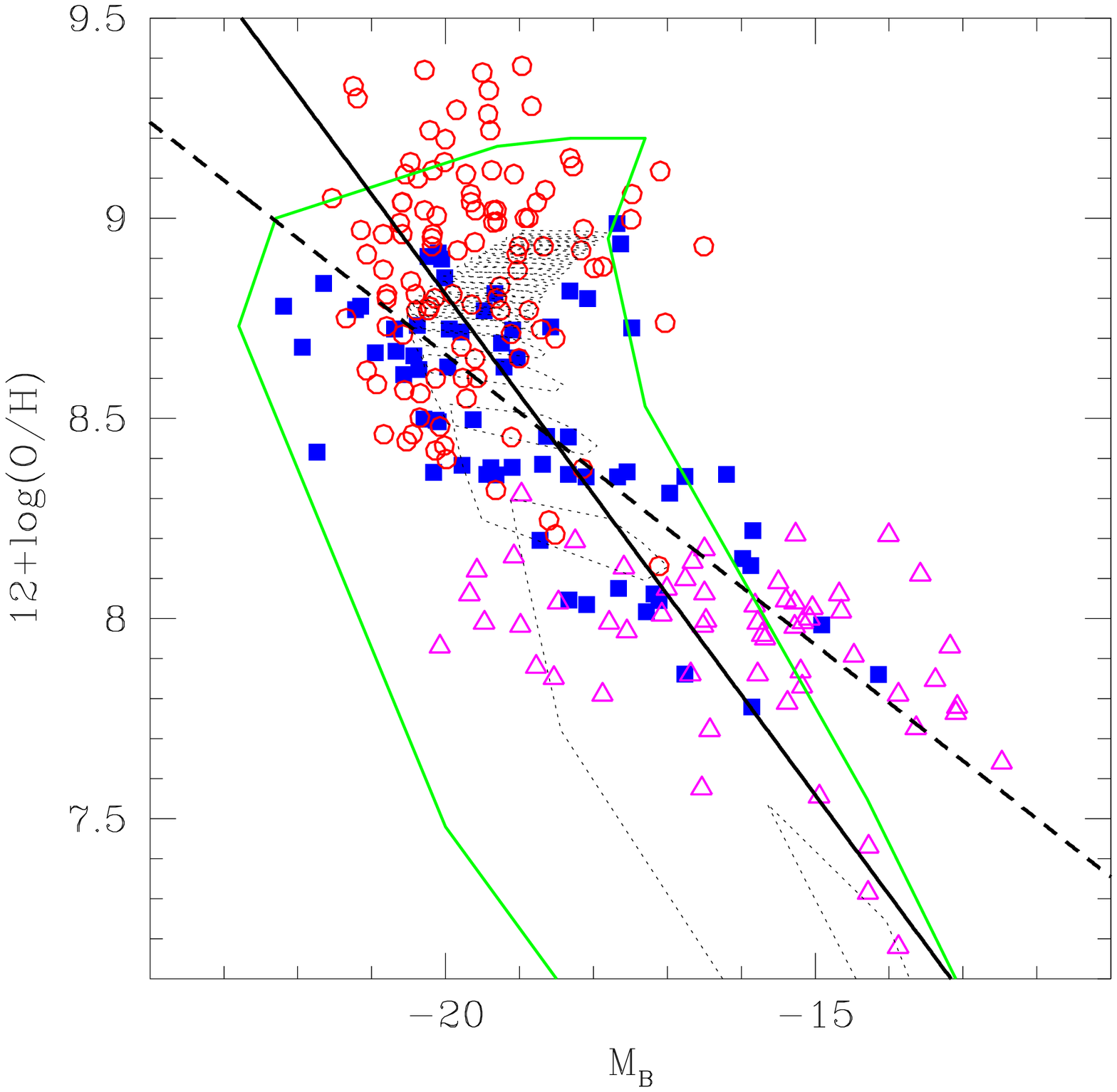}
\caption{{\it Left}: N/O versus \doh\ for the samples of star-forming galaxies 
(dots) described in Figure~\ref{UV}. Model predictions assuming 
bursting star formation scenario are shown. The model parameters are listed 
in Table 1 of \cite{MC01}. 
This figure shows the dichotomy between 
the models which reproduce the scatter in the metal-poor region 
(\doh\ $\le 8.5$), and those which reproduce the scatter 
in the metal-rich region (\doh\ $\ge 8.5$).
{\it Right}: The metallicity-luminosity relation for the starburst galaxy 
samples described in Figure~\ref{UV}: \hii\ galaxies (triangles), SBNGs 
(circles) and UV-selected galaxies (squares). The continuous thick line 
is a linear fit to the whole sample of starburst galaxies, and 
the dashed thick line is a linear fit to local {\it ``normal''} irregular 
and spiral galaxies (\cite{KZ99}). Model predictions assuming
a bursting star formation scenario are shown. The model parameters are
the same as those used in the left panel. The delineated area 
(thick green line) encompasses all the model predictions calculated 
with the parameters reported in Table 1 of \cite{MC01}. For
illustration, the prediction of a specific model (dotted thin line) 
is shown.}
\label{models}
\end{figure}

At a given metallicity, the distribution of N/O abundance ratios 
shows a large dispersion, both at low and high metallicity 
(\cite{Cozetal99,Contetal01}). 
Only part of this scatter is due to uncertainties in the abundance 
determinations. The additional dispersion must therefore be 
accounted for by galaxy evolution models.
Various hypotheses (localized chemical ``pollution'', IMF variations, 
differential mass loss, etc) were suggested as responsible for such a
scatter (e.g.\,\cite{KS98}) but none of them is able to 
reproduce the full N/O scatter at a given metallicity. 

A natural explanation for the variation of N/O at constant metallicity 
might be a significant time delay between the release, into the ISM, 
of oxygen by massive, short-lived stars and that of nitrogen produced 
in low-mass longer-lived stars (see Figure~\ref{UV}b). 
The ``delayed-release'' model assumes that star formation is an 
intermittent process in galaxies, while maintaining an universal IMF 
and standard stellar nucleosynthesis (\cite{EP78,G90}). 

\cite{MC01} recently investigated this possibility in order to quantifiy 
the star formation history of starburst galaxies. The observed
dispersion in the well-known metallicity-luminosity relation 
has been used as an additional constraint. 
It was confirmed that {\em continuous} star formation models are 
unable to reproduce the scatter observed in both N/O and \mabs\ 
versus O/H scaling relations. 
The dispersion associated with the distribution of N/O as a function 
of metallicity can indeed be explained in the framework of {\em bursting} 
star formation models. Figure~\ref{models} shows the oscillating behavior 
of the N/O ratio due to the alternating bursting and quiescent phases. 
In this case, the observed dispersion in the N/O versus O/H relation is 
explained by the time delay between the release of oxygen by massive 
stars into the ISM and that of nitrogen by intermediate-mass stars 
(see also Figure~\ref{UV}b for a schematic view). 
During the starburst events, as massive stars dominate the chemical 
enrichment, the galaxy moves towards the lower right part of the
diagram. During the interburst period, when no star formation is 
occurring, the release of N by low and intermediate-mass stars occurs 
a few hundred Myrs after the end of the burst and increases N/O at 
constant O/H. The dilution of interstellar gas by the newly accreted 
intergalactic gas is also observed during the quiescent phases. 

Extensive model computations (see \cite{MC01} for details) 
show that no possible 
combination of the model parameters (i.e., burst duration, interburst period, 
star formation efficiency, gas accretion timescale, etc) is able to account 
for the observed spread for the whole sample of galaxies. 
\cite{MC01} found that the most important parameter for 
reproducing the observed spread is the star formation efficiency. 
Once the star formation efficiency is set, the extent of the predicted 
spread mainly depends on the starburst duration. The models show that, to 
account for the observed scatter of N/O for the whole metallicity range, 
one needs to consider at least two {\em regimes} (low and high metallicity) 
characterized by different star formation efficiencies and starburst durations.

Following these models, metal-rich (\doh\ $\ge 8.5$) spiral galaxies 
differ from metal-poor ones by a higher star formation 
efficiency and starburst frequency. Moreover, the fit is very good 
for metal-rich galaxies if the quiescent period between two 
successive bursts is of the same order as the gas infall 
timescale. This can be understood 
in the context of the hierarchical galaxy formation scenario where 
a major burst of star formation is activated each time a galaxy 
undergoes a minor merger event, such as the accretion of a small 
satellite or primordial HI gas clouds (e.g.\,\cite{Coleetal00}). 

Finally, these models confirm previous 
claims (\cite{Contetal01}) that UV-selected galaxies are 
observed at a special stage in their evolution. Their low N/O 
abundance ratios with respect to other starburst galaxies is
well explained if they have just undergone a powerful starburst 
which enriched their ISM in oxygen.

\vspace{-.2cm}
\section{Chemical properties of intermediate-- and high--redshift galaxies}
\vspace{-.3cm}

\begin{quote}
\begin{figure}
\plottwo{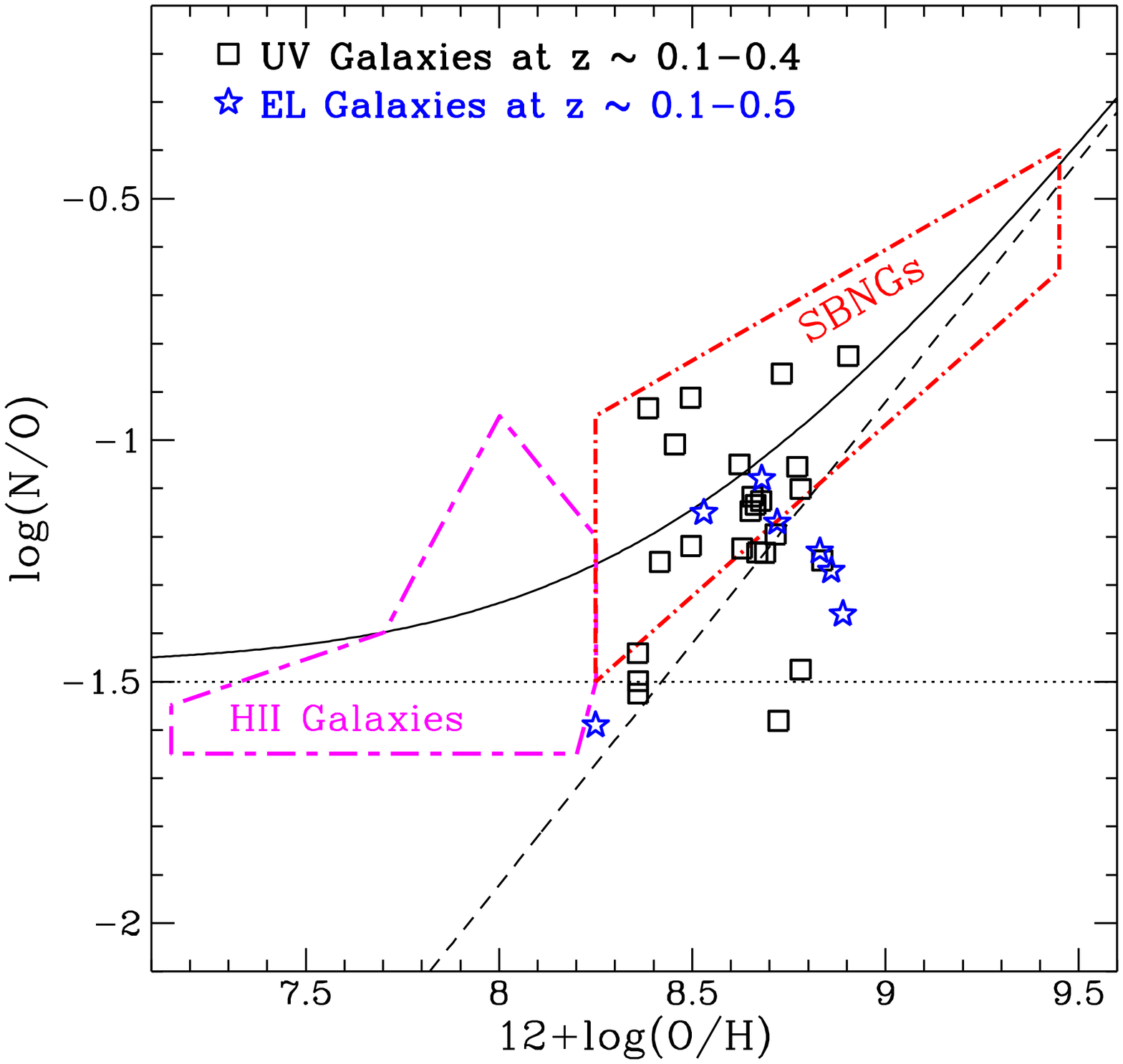}{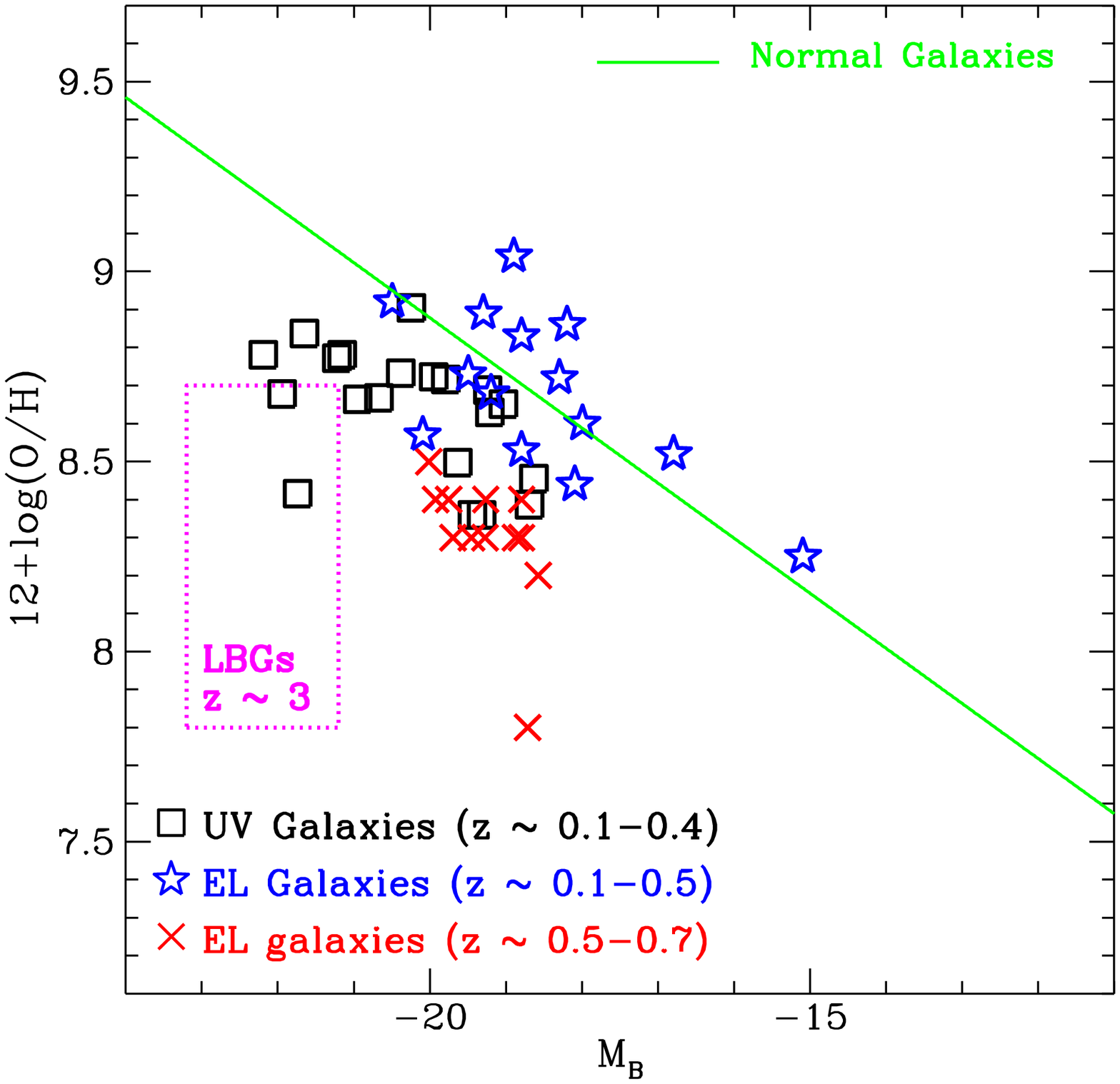}
\caption{N/O ({\it left}) and absolute $B$-band magnitude ({\it right}) as 
a function of metallicity for intermediate and high redshift star-forming 
galaxies. All chemical abundance data published so far for distant galaxies 
are on this plot. The samples of intermediate-redshift galaxies are: 
UV-selected galaxies (squares; \cite{Contetal01}) and emission-line 
(EL) galaxies (stars; \cite{KZ99}) at $z \sim 0.1-0.5$, 
compact and luminous EL galaxies at $z \sim 0.5-0.7$ (crosses; 
\cite{Hametal01}). 
The location of high-redshift ($z \sim 3$) Lyman break galaxies is shown as 
a box encompassing the range of O/H and \mabs\ derived for these objects 
(\cite{Petetal01}). On the left panel, the location of nearby \hii\ 
galaxies and SBNGs (see Figure~\ref{UV} for references) is shown for 
comparison. The metallicity--luminosity relation for local ``normal'' galaxies 
(see \cite{KZ99}) is shown as a solid line on the right panel.}
\label{highz}
\end{figure}
\end{quote}

In Figure~\ref{highz} ({\it left}), we compare the location of 
intermediate-redshift ($z \sim 0.1-0.4$) emission-line 
(EL; \cite{KZ99}) and UV-selected 
(\cite{Contetal01}) galaxies with nearby 
samples of \hii\ and SBNGs in the N/O versus O/H plane. 
Most of these distant objects have chemical abundances typical of massive and 
metal-rich SBNGs (i.e., \doh\ $\geq 8.5$). Some of them show high 
N/O abundance ratios (log(N/O) $\ga -1$) which could be due to 
a succession of starbursts over the last few Gyrs (e.g.\,\cite{Cozetal99}). 
None of these intermediate-redshift galaxies 
are located in the region occupied by nearby metal-poor \hii\ galaxies.
The fact that intermediate-redshift galaxies are mostly metal-rich 
objects is likely to be a selection effect arising 
from the well-known metallicity--luminosity relationship. Only the 
most luminous, and thus metal-rich objects, are detected so far with 
the current instrumentation on 8-m class telescopes. It is interesting to note 
that, like a significant fraction of UV-selected galaxies, some 
optically-selected EL galaxies also show strikingly low N/O ratios.

The location of distant galaxies in the metallicity--luminosity relation 
is shown in Figure~\ref{highz} ({\it right}). Three samples of
intermediate-redshift galaxies are considered: EL galaxies at 
$z \sim 0.1-0.5$ (\cite{KZ99}), luminous and 
compact EL galaxies at $z \sim 0.5-0.7$ (\cite{Hametal01}) and 
the UV-selected galaxies (\cite{Contetal01}). 
The location of high-redshift ($z
\sim 3$) Lyman break galaxies (LBGs) is shown as a box encompassing the
range of O/H and \mabs\ derived for these objects (\cite{Petetal01}). 
Whereas EL galaxies with redshifts between $0.1$ and $0.5$
seem to follow the metallicity--luminosity relation of ``normal'' 
galaxies (solid line), there is a clear deviation for both UV-selected 
and luminous EL galaxies at higher redshift ($z \sim 0.5-0.7$). 
These galaxies thus appear $2-3$ mag brighter than
``normal'' galaxies of similar metallicity, as might be expected
if a strong starburst had temporarily lowered their mass-to-light
ratios. \cite{Hametal01} argue that luminous and compact EL galaxies 
could be the progenitors of present-day spiral
bulges. The deviation is even stronger for LBGs at $z \sim 3$. 
Even allowing for uncertainties in the determination 
of O/H and \mabs, LBGs fall well below the 
metallicity--luminosity relation of ``normal'' local galaxies and 
have much lower abundances than expected from this relation given 
their luminosities. The most obvious interpretation (\cite{Petetal01}) 
is that LBGs have mass-to-light ratios significantly lower than those 
of present-day ``normal'' galaxies. 

Although the samples are still too small to derive firm conclusions on 
the link between distant objects and present-day galaxies, the present 
results give new insight into the nature and evolution of distant 
star-forming galaxies. No doubt that the next large-scale spectroscopic 
surveys on 10-m class telescopes (e.g.\,VIRMOS on VLT, COSMOS/EMIR on 
Grantecan, etc) will shed light on these fundamental issues by producing 
statistically significant samples of galaxies over a large range of redshifts.

\acknowledgements{We warmly thank Laurence Tresse and Marie Treyer, 
the organizers of this wunderbar conference together 
with all the people involved in the LOC (fifi, seb, henry, etc).}

\vfill
\end{document}